# DEVELOPMENT OF ULTRA-THIN POLYETHYLENE BALLOONS FOR HIGH ALTITUDE RESEARCH UPTO MESOSPHERE

B. Suneel Kumar*[†], N. Nagendra[†], D. K. Ojha[†¶], G. Stalin Peter[†], R. Vasudevan[†], D. Anand[†], P. M. Kulkarni[†], V. Anmi Reddy[†], T. V. Rao[†], and S. Sreenivasan[†]

[†]*Tata Institute of Fundamental Research, Balloon Facility, P.B no: 5, ECIL post office, Hyderabad-500 062, India.*

[†¶]*Tata Institute of Fundamental Research, Homi Bhabha Road, Colaba, Mumbai -400 005, India.*

*Corresponding Author:* B. Suneel Kumar, E-mail: bsuneel@tifr.res.in

**Abstract**

Ever since its inception four decades back, Balloon Facility of Tata Institute of Fundamental Research (TIFR), Hyderabad has been functioning with the needs of its user scientists at its focus. During the early nineties, when the X-ray astronomy group at TIFR expressed the need for balloons capable of carrying the X-ray telescopes to altitudes up to 42 km, the balloon group initiated research and development work on indigenous balloon grade films in various thickness not only for the main experiment but also in parallel, took up the development of thin films in thickness range 5 to 6 µm for fabrication of sounding balloons required for probing the stratosphere up to 42 km as the regular 2000 grams rubber balloon ascents could not reach altitudes higher than 38 km. By the year 1999, total indigenisation of sounding balloon manufacture was accomplished. The work on balloon grade ultra-thin polyethylene film in thickness range 2.8 to 3.8 µm for fabrication of balloons capable of penetrating mesosphere to meet the needs of user scientists working in the area of atmospheric dynamics commenced in 2011. Pursuant to the successful trials with 61,000 cu.m balloon made of 3.8 µm Antrix film reaching stratopause (48 km) for the first time in the history of balloon facility in the year 2012, fine tuning of launch parameters like percentage free lift was carried out to take the same volume balloons to higher mesospheric altitudes. Three successful flights with a total suspended load of 10 kg using 61,000 cu.m balloons were carried out in the month of January 2014 and all the three balloons crossed in to the mesosphere reaching altitudes of over 51 km. All the balloons flown so far are closed system with no escape ducts. Balloon fabrication, development of launch hardware, flight control instruments and launch technique for these mesospheric balloon flights are discussed in this paper.

## 1. Introduction

A complete understanding of the stratospheric wind pattern over different seasons in general as well as latest upper stratospheric wind data in particular are the two essential pre-requisites before launch of balloons from TIFR Balloon Facility, Hyderabad (latitude: $17.47^0$ N, longitude: $78.58^0$ E), India. The main reason for this requirement is the asymmetric distribution of the flight operational area around the balloon base. The best terrain for safe recovery of the payload is located west of the base station covering the entire Deccan plateau extending up to 486 km, while the dense reserve forests and mountains in the north, east and south limit the operational range to 396 km, 364 km and 392 km respectively, despite allotment of a larger operational area shown in Fig. 1. It is therefore essential to know the expected flight trajectory beforehand to ensure that the user scientists get the required float duration. The stratospheric wind up to 42 km altitude over Hyderabad, during flight season (October to April) was analysed by using United Kingdom Meteorological Office stratospheric assimilation data at different altitudes (Manchanda *et al.*, 2011).

For this purpose, stratospheric wind probing flights are conducted before the main flight for predicting the flight path and measurement of essential atmospheric parameters like temperature, humidity and pressure. Typically, rubber balloons weighing up to 2000 grams, carrying radar targets were in use for past several decades and recently global positioning system (GPS)-sondes are used for providing the wind and meteorological data up to 38 km. The rubber balloons served the purpose till 1989, as all the scientific balloon flights were below 38 km altitude. In anticipation of generating wind data beyond 38 km, the plastic sounding balloons of volume 4,364 cu.m fabricated out of 5.8 µm film for probing upper wind data up to 42 km, were imported as a stop gap measure. Simultaneously, efforts were made to manufacture the plastic sounding balloons indigenously and as a first step, the raw material for the balloon production viz. thin polyethylene film (5.8 µm) was imported from abroad and the trial production of the sounding balloons commenced in 1991.

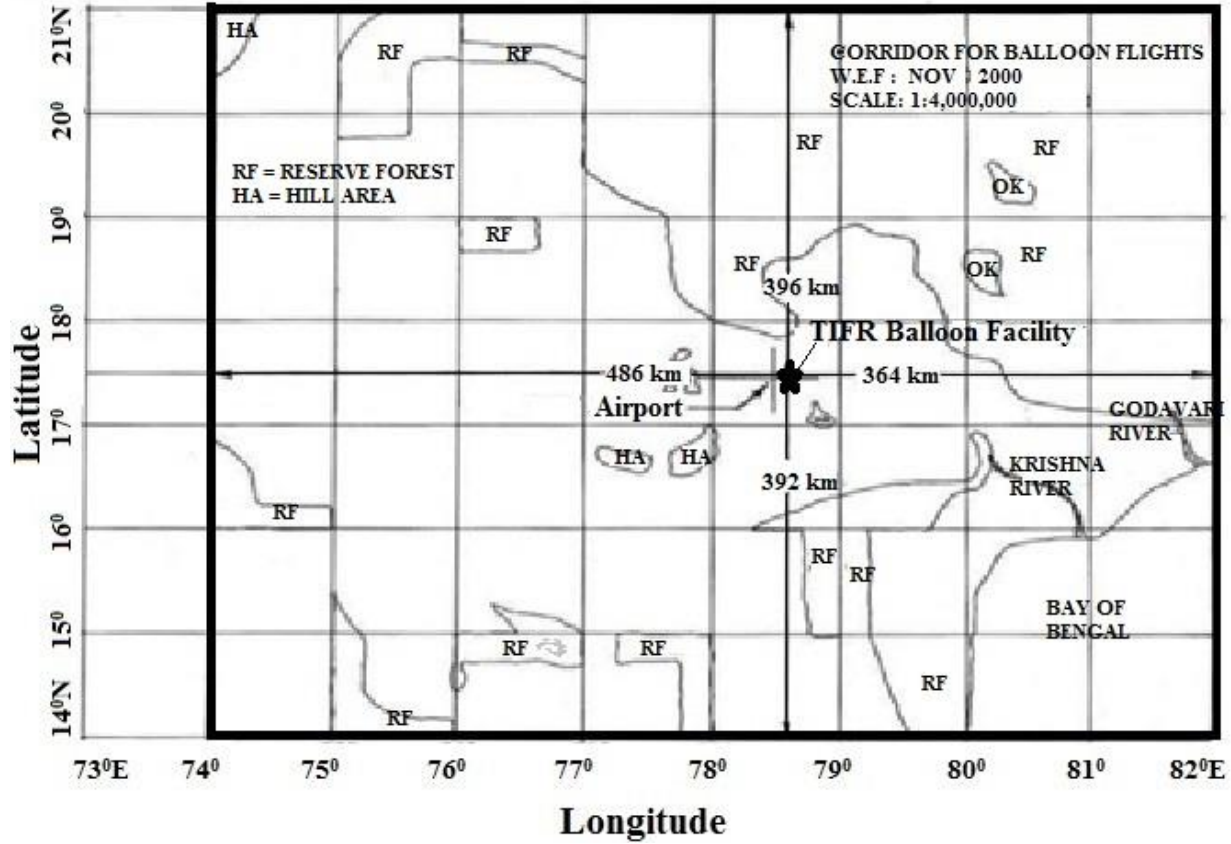

**Fig.1.** Balloon operational corridor map marked with bold solid lines.

In 1993, user scientists of X-ray astronomy group from TIFR came up with demand for flights capable of reaching up to 42 km and to meet this requirement TIFR balloon facility augmented its balloon manufacturing capability in 1994. In the meanwhile, the indigenously manufactured sounding balloons proved to be as good as the imported ones and hence as the next step, it was decided to extrude the thin film locally so that the indigenisation process is complete. The mechanical properties of ANTRIX balloon film developed by the TIFR balloon facility and fabrication of single cap large volume balloons by using this film for different scientific experiments have been discussed elsewhere (Suneel Kumar *et al.,* 2008).

In 1998, trial extrusion of thin polyethylene film of thickness 6 µm was carried out successfully and based on promising laboratory test results, one ton of 6 µm film was extruded in the year 1999 for fabrication of sounding balloons. The balloons successfully reached altitudes up to 42 km and provided valuable wind data. During the 3 ton extrusion carried out in 2006, the film thickness was further reduced to 5.8 µm leading to significant reduction in balloon weight and the sounding balloons fabricated using this film performed well, the details of which are discussed in Manchanda *et al.,* (2006).

## 2. Development of Ultra-thin Balloon Film

Over the years, scientists began to show interest in the study of atmospheric dynamics and meteorological parameters over middle atmospheric region up to mesosphere (~50 km) using balloon-borne experiments. The study involves balloons with a payload carrying capability of about 25 kg reaching mesosphere. This has led to the development of the basic raw material namely ultra-thin polyethylene film with thickness ranging from 3.8 µm to 2.8 µm at TIFR balloon facility, Hyderabad.

In order to reach mesosphere (~50 km) altitude, the following requirements have to be satisfied.

1. The balloon film must be ultra-thin to minimize weight compared to conventional scientific plastic balloons.

2. The ultra-thin film should have the high tensile strength to withstand large stresses during ground bubble inflation.

3. The balloon flight housekeeping electronics such as telemetry, telecommand, timer, GPS, etc. must be light weight and consume very low power.

The first attempt in developing the ultra-thin films and launching of these scientific balloons were made by Japanese balloon group of the Institute of Space and Astronautical Science (ISAS), Japan. In 2000, they launched a light weight (6.8 kg) balloon of volume 5,000 cu.m and it successfully reached an altitude of 43 km (Saito *et al.*, 2002). Extending their work in 2004, they fabricated and launched a balloon of volume 30,000 cu.m with 3 µm film thickness and it reached an altitude of 50.7 km, whose details are discussed in Saito *et al.*, (2006). The prerequisite for developing such a balloon is the mechanical properties of the balloon film used. The ultra-thin film made out of linear low density polyethylene (m-LLDPE) resins with added metallocene catalysts proves to have outstanding toughness with improved tensile strength, impact resistance and puncture performance. The material properties of ultra-thin film are discussed elsewhere in Yamagami *et al.*, (2004). The fabrication process of the balloon with these materials also has the advantage of lower heat-seal initiation temperatures and higher hot-tack strength for faster line speeds with excellent seal integrity.

### 2.1. *First trial extrusion of ultra-thin balloon film*

Based on this valuable input, trial extrusion was performed in the year 2008 for extruding ultra-thin film using m-LLDPE resin in a commercial plastic extrusion factory. No special efforts were made to augment the extruder for the production of thin film. We extruded 200 kg of 3.8 µm lay flat tubing with 1300 mm web width by gradually lowering the gauge thickness from 13 µm to 3.8 µm. During extrusion, some problems like frequent bubble breaks in between and irregular web width were faced due to non-stability of extrusion bubble. Laboratory (tensile) tests at room temperature gave us encouraging results qualifying the film as a candidate material for balloon fabrication.

### 2.2. *Balloon fabrication using 3.8 µm ultra-thin film*

One trial balloon (serial no: TS-16111) with 3.8 µm ultra-thin film was fabricated with the balloon design remaining the same as regular sounding balloon of 4,077 cu.m volume with 28 gores. The balloon was sealed on a curved table with only buffer strips of thickness 20 µm and width 33 mm. The balloon weighed about 7 kg which is less by 2 kg compared to a similar volume balloon made out of 5.8 µm. This trial balloon was launched with a GPS-sonde (EN-SCI make) of 740 gm weight, inflated with a gross lift of 10.65 kg which includes a free lift of 2.9 kg (37.4% of total load) on 2011 December 1 at

22:10 hrs (IST). The balloon ascended with an average ascent rate of 284 meters per min and reached a maximum altitude of 44.05 km (a regular sounding balloon reaches a maximum altitude of 43 km) in 2 hrs 35 minutes. This was the first time a TIFR balloon reached a record altitude in India. Regarding the performance of the balloon, it was normal during ascent and successfully crossed the coldest tropopause and after reaching the ceiling altitude, the balloon initially descended and started floating at the same altitude for few minutes instead of bursting. Normally, sounding balloons made of regular LLDPE resins burst on their own after reaching the ceiling altitude, due to build-up of super pressure, and then descend down to ground. The configuration and time-height curve of this flight are shown in Fig. 2.

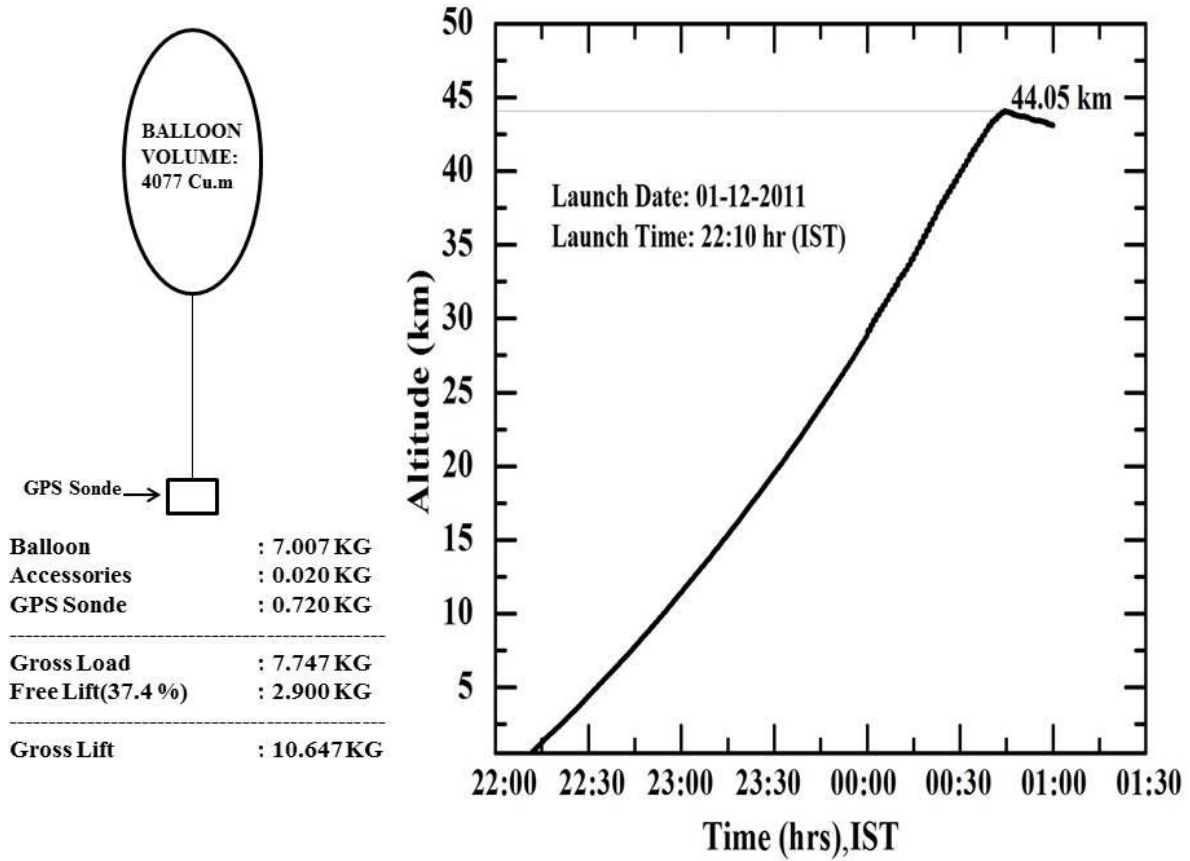

**Fig.2.** The flight configuration and time-height curve for TS-16111 balloon.

After observing this abnormal behaviour of ultra-thin film balloon, we inferred that the balloon instead of bursting might have developed a small hole due to super pressure. This could have provided a very small venting area for escape of entire lift gas (hydrogen). The balloon finally descended after several hours of float. After reviewing the above phenomenon, balloon facility launch crew decided to develop a hole at the top portion of the balloon after it attains its maximum altitude by fusing and tearing the balloon film by heating. The electronics and communication group at balloon facility developed a micro-controller based mini-timer for fusing the balloon top portion using Ni-chrome wire. The mini-timer weighing around 250 grams operates on 6V battery. For validating this concept, a test flight was conducted on 2011 December 29 using a regular sounding balloon of volume 4,077 cu.m made out of 5.8 μm film with mini-timer package incorporated at the top apex and the 100 mm Ni-chrome wire attached to top portion of balloon. The mini-timer was intentionally set to operate at 130 min after launch. The balloon ascended normally with an average ascent rate of 250 meters per min. The balloon could reach up only to an altitude of 40.4 km in 130 min and started descending. It is inferred from this observation that the mini-timer operated before the balloon could reach its maximum altitude and created a hole at the top portion during balloon ascent phase. Fig. 3 shows the flight configuration and complete time-height curve. Based on the successful performance of the mini-timer and Ni-chrome wire flight termination system, it was decided to use mini-timer for balloon termination for all sounding as well as balloons made using ultra-thin films.

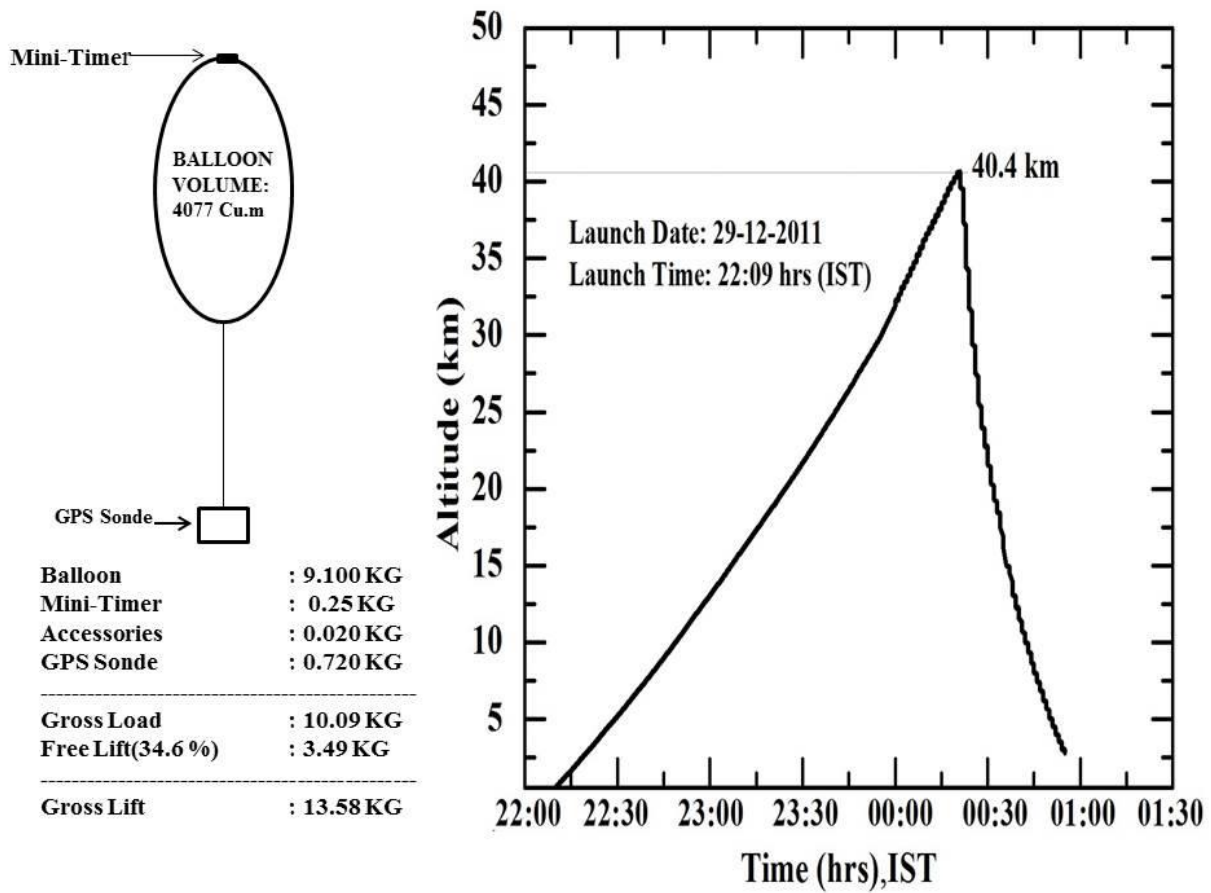

**Fig. 3.** The flight configuration and complete time–height curve for the sounding balloon made with 5.8 μm.

After the successful performance of the trial balloon, extrusion of 3.8 μm ultra-thin balloon film was successfully taken up during February 2012. Along with regular production we also extruded small quantity of 3.4 μm and 3.2 μm thin film for trial purpose and conducted tensile tests on tensile testing machine at balloon facility, Hyderabad at both room temperature and at -80$^0$C. Table 1 shows the test results in machine direction (MD) and transverse direction (TD). The test results show that the film is suitable for regular production of ultra-thin film balloons.

Table 1. Tensile test for the quality checking of ultra-thin film at different temperatures.

| Temperature (°C) | Thickness (μm) | Width (cm) | Ultimate strength (kg/cm$^2$) | | Ultimate elongation (%) | |
|---|---|---|---|---|---|---|
| | | | machine direction (MD) | transverse direction (TD) | machine direction (MD) | transverse direction (TD) |
| +23°C | 3.8 | 132 | 530 | 440 | 490 | 880 |
| | 3.4 | 131 | 470 | 400 | 380 | 940 |
| | 3.2 | 134 | 600 | 380 | 430 | 815 |
| -80°C | 3.2 | 134 | 765 | 360 | 160 | 90 |

### 2.3. *Trial balloon fabrication using 3.4 μm film*

A second trial balloon of volume 5,014 cu.m (serial no: TS-16212) was fabricated using 3.4 μm film. This balloon despite being slightly larger in volume than the regular sounding balloon weighed only 6.3 kg and lighter than first trial balloon by 0.7 kg. A test flight was conducted on 2012 March 8 using this balloon with a suspended payload of 0.72 kg. A free lift of 2.8 kg (39.77% of the total load) was added to the gross load of 7.3 kg. The mini-timer package incorporated at the top apex and loaded the pre-set time as 02:15 hrs (IST) on 2012 March 9. The balloon was launched at 22:55 hrs (IST) and ascended with an average ascent rate of 276 meters per min and reached a maximum altitude of 45.03 km in 2 hrs 43 minutes at 01:38 hrs (IST). After reaching the maximum altitude the balloon started descending to ground on its own and mini-timer operated at pre-set time during descent. The performance of balloon was normal during all phases. The flight configuration and time-height curve are shown in Fig. 4. Two more balloons of volume 5,014 cu.m were fabricated using 3.2 μm thin film with the extraneous portion of seal trimmed for further reduction in weight. The balloons weighed 5.97 kg each (serial no's: TS-16312 & TS-16412).

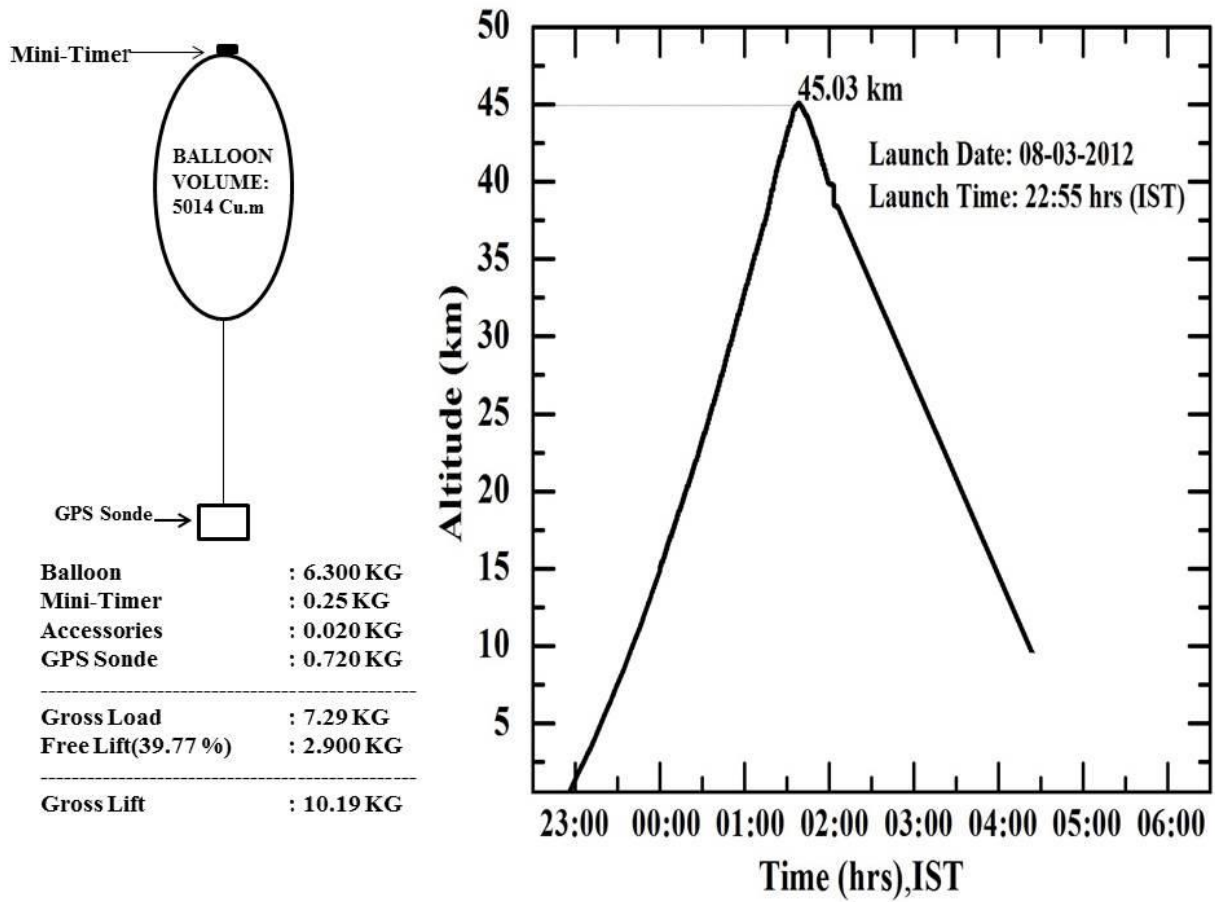

**Fig. 4.** The flight configuration and complete time-height curve for TS-16212 balloon.

## 3. Design and Fabrication of 61,000 cu.m Balloon

After reaching 45.03 km altitude using 5,014 cu.m balloon in previous attempts, it was decided to make a balloon capable of flying 10 kg payload to 50 km. After several iterations, the volume of the balloon was calculated as 61,013 cu.m with 3.8 µm film as shell thickness for reaching an altitude of 52 km. The gore length for this balloon worked out to be 74.16 meters with 69 gores. The designed balloon was a closed system and does not have any venting ducts at the base. Before taking up the fabrication of the 61,013 cu.m balloon, a trial balloon of 11 m length was fabricated using 3.8 µm film with 6 gores, to test destruct device operation, inflation through top apex and the ease with which the balloon can be handled. The top portion was collapsed for incorporating 150 mm diameter nylon ring with 20 m filling tube attached to it. An acrylic destruct device button was also incorporated at 3 m from top apex. The trial balloon was inflated with hydrogen gas (approximately a total of 50 cu.m hydrogen was used) till deployment of all gores. During inflation, it was observed that the look of the material was apparently delicate but there were no marks of yielding and the filling operation was found normal. The inflated balloon was raised to 5 m from the ground and a dead weight of 10 kg was attached to the destruct device rope. The hydrogen gas escaped as the destruct device successfully tore out the rip panel. Several problems were encountered during fabrication of this balloon. While sealing the balloon, because of straight table, thin film sealing was not possible as the film was touching the heater bar and getting damaged. To perform flawless sealing, balloon group fabricated a curved wooden table with the same shape as balloon and developed a continuous band sealing machine with drive. For reducing balloon weight a buffer strip of 25 mm (width) x 20 µm (thickness) was used. Fig. 5 shows the sealing machine and curved table.

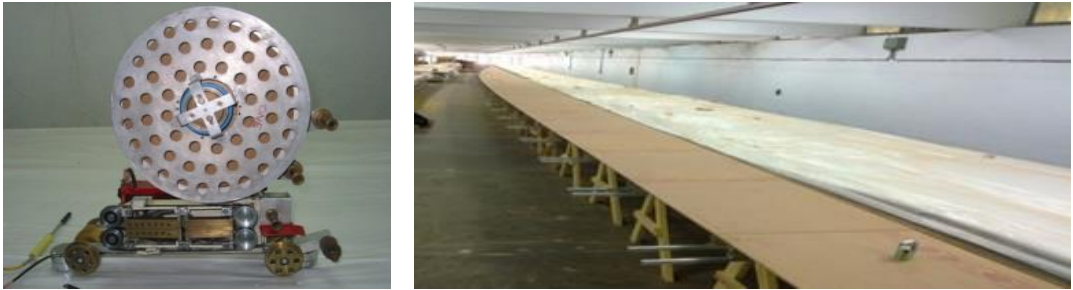

**Fig. 5.** The sealing machine (left) and curved table (right) used in balloon fabrication.

Four such balloons have been fabricated with weights ranging from 34.5 kg to 36 kg on curved table using self-driven continuous band sealing machine. The difference in weight could be attributed to non-linearity in gauge thickness of film. A destruct device was incorporated to all the balloons as an additional safety to bring down the payload in case the balloon does not burst after attaining super pressure at ceiling altitude. One filling tube was attached at the top apex ring. As per the design, this balloon could carry a maximum payload of 10 kg including the main scientific instrument, flight control packages, load line, cables, air traffic control (ATC) transponder, GPS-sonde and a small parachute for safe recovery. All the four balloons were fabricated using 3.8 µm film which was extruded during initial trial extrusion in 2008. The details of balloons are shown in Table 2.

**Table 2.** List of balloons fabricated with 3.8 µm film (2008 lot).

| S.No. | Balloon serial number | Weight (kg) |
|---|---|---|
| 1 | MS-16512 | 34.8 |
| 2 | MS-16612 | 35.2 |
| 3 | MS-16712 | 36.0 |
| 4 | MS-16812 | 34.5 |

### 3.1. *Launching of 61,013 cu.m balloon*

To achieve higher altitudes, the control instrumentation packages must be very compact and light weight. TIFR electronics and communication group designed and developed a single card command integrated with timer, global system for mobile communications (GSM)-GPS unit and a mini-timer. The single card command and timer system is designed for controlling balloon flight with small payloads and can provide limited commands like main cut, time extension, and timer cut operations and one command for package health check. A 3 watt radio data modem operating at 148.25 MHz (TSLM) is used to telemeter the status of commands and other health parameters at 9600 baud rate. On payload touch down, the GSM-GPS unit sends its physical position continuously as text message to configured mobile phone or to the tracking server every minute. The mini-timer is used to terminate the flight at a pre-determined time. The total weight of the package was 3 kg. The single card command, GSM-GPS, mini-timer and radio data modem are shown in Fig. 6.

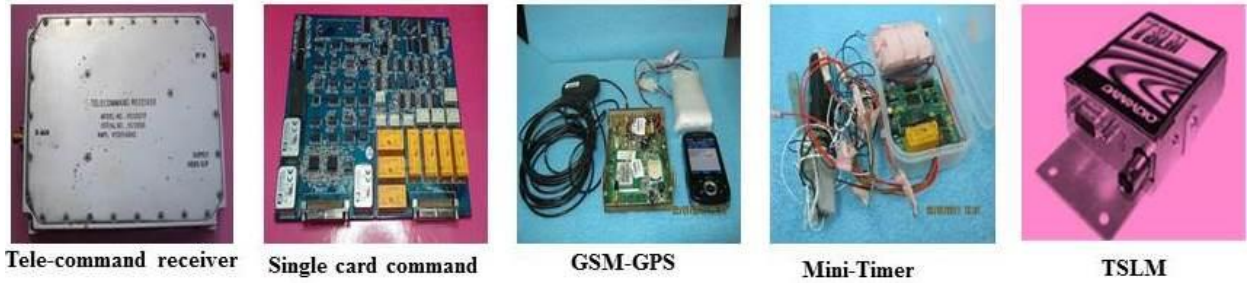

**Fig. 6.** Control instrumentation packages used in balloon-borne experiments.

### 3.2. *Mechanical launcher*

For launching regular sounding balloons made out of 5.8 μm thin film, we adopted hand launch method, where a person physically holds the balloon bubble after inflation, as the gross lift is of the order of only ~ 10 kg. However, the gross lift in case of 61,013 cu.m balloon worked out to 59 kg including free lift and hence it is not possible for one person to hold and restrain the bubble from lifting off. Further, it is not advisable to handle ultra-thin film balloons in this manner which leads to film damage. Hence, the TIFR balloon facility workshop group designed and developed a mechanical launcher after getting valuable inputs from the Japanese balloon group. In 1993, ISAS developed a new launch gear for high altitude balloon-borne experiments and made successful balloon launches by their newly developed launch gear (Yamagami *et al.,* 1998). The mechanical launcher is a simple device comprising spring loaded aluminium clamps for holding the bubble during inflation. An air pillow was made with soft rubber to hold balloon material covered with soft flannel cloth. A pump was used for inflating air into pillow. Fig. 7 shows the mechanical launcher and the balloon passing through the mechanical launcher.

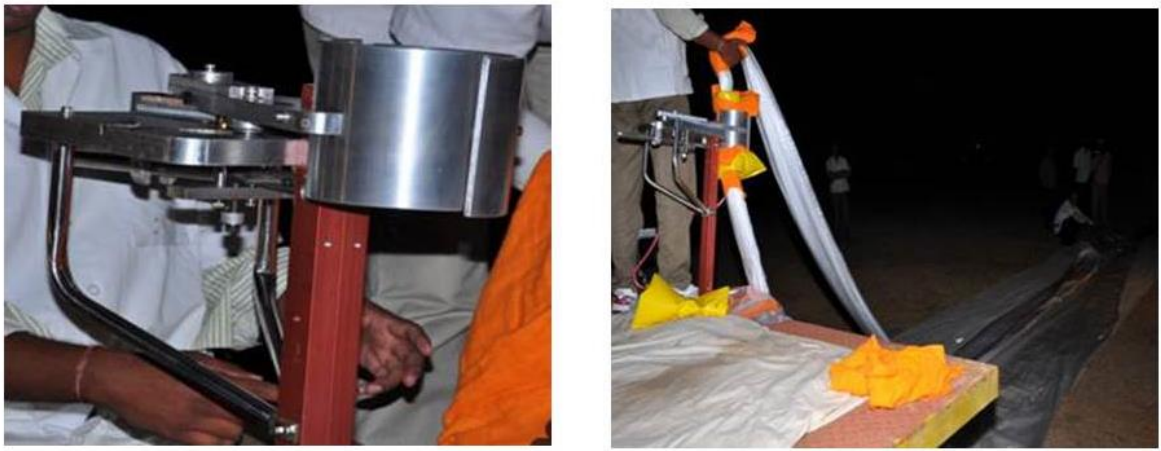

**Fig. 7.** A close-up view of mechanical launcher (left) and a view of the balloon passing through mechanical launcher (right).

### 3.3. *Launching first 61,013 cu.m balloon*

On 2012 April 25, the first balloon (serial no: MS-16712) was launched carrying 9.7 kg total suspended weight. The launch was successful and the mechanical launcher performed well and the payload was released by hand. The gross lift of 54.8 kg hydrogen gas was filled in balloon using pressure and temperature (P&T) measurement method and completed in 12 minutes. A small 4m diameter parachute was used for safe descent of control instrumentation packages. We used integrated package comprising single card command, TSLM, mini-timer, ATC transponder, GSM-GPS unit and an independent GPS-sonde unit for position and altitude information. An auxiliary balloon was attached to the suspended load to assist smooth take-off. Dynamic launch method was employed for payload release. The balloon was launched at 03:25 hrs (IST) and after 3 minutes, first command was given to separate the auxiliary launch balloon. The balloon ascending at an average rate of 235 meters per min passed through a high relative humidity region leading to condensation on the GPS-sonde's polystyrene box due to which the GPS antenna gain was grossly diminished. The GPS gave the position information intermittently and finally lost 3D fix at an altitude of 39.7 km. However, GPS-sonde gave altitude data based on the pressure and temperature sensors. Based on this altitude data, the balloon ascended up to 48.0 km at 06:50 hrs (IST) and started floating at the same altitude. Flight was terminated by sending main cut command at 08:00 hrs (IST). The payload descended with parachute and landed 100 km west of launching site. This balloon launched under high altitude balloon development (HAA) project and designated as HAA-01 flight reached the stratopause (~48 km) for the first time from India. Fig. 8 shows the flight configuration and time-height curve.

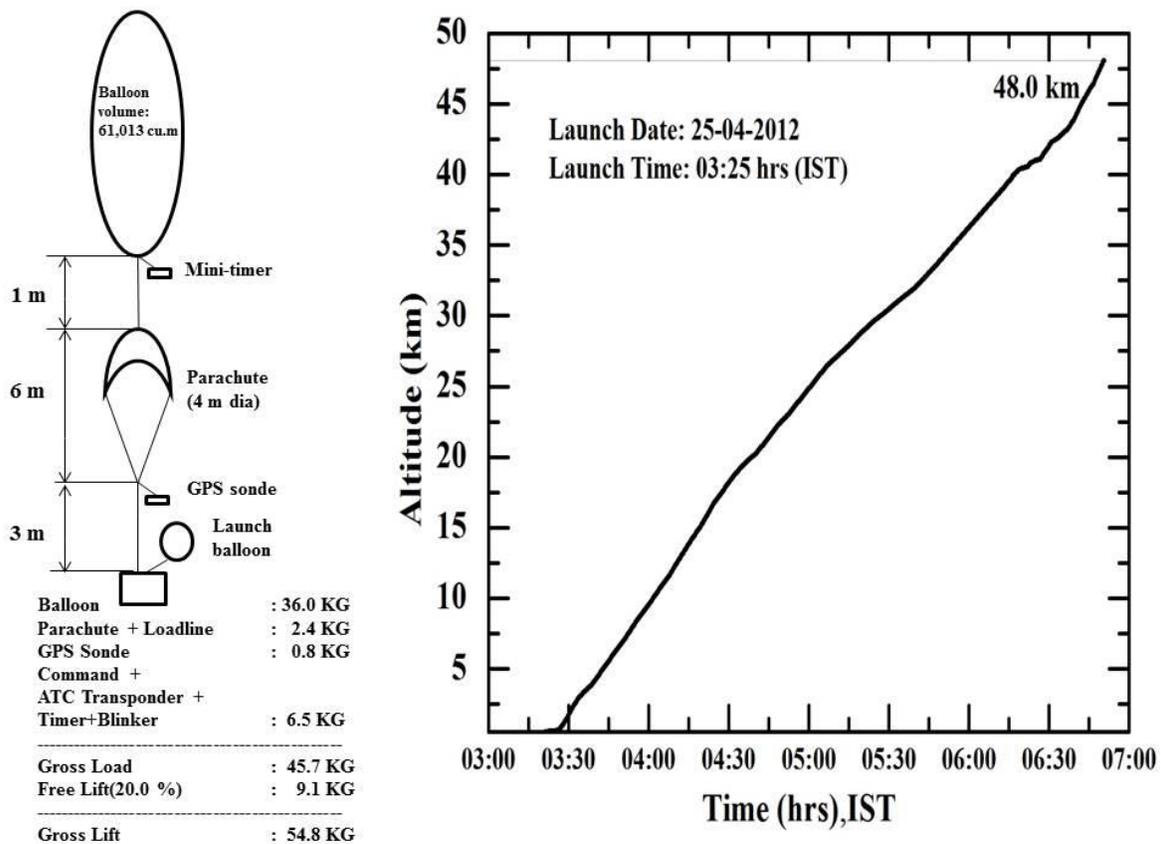

**Fig. 8.** The flight configuration and time-height curve for HAA-01 balloon.

GPS-sonde is a crucial instrument that provides the altitude data. Therefore its thermal packing was improvised, to overcome the problem faced in HAA-01 flight, and to test its performance a test flight was conducted on 2012 May 17 using 5,014 cu.m thin film balloon (serial no: TS-16312). The balloon was launched at 22:00 hrs (IST) and it reached a maximum altitude of 47.0 km in 2 hrs 50 minutes with an average ascent rate of 276 meters per min and burst at 00:50 hrs (IST). The GPS-sonde functioned very well throughout the flight both in ascent and descent phases. The flight configuration and time-height curve are shown in Fig. 9.

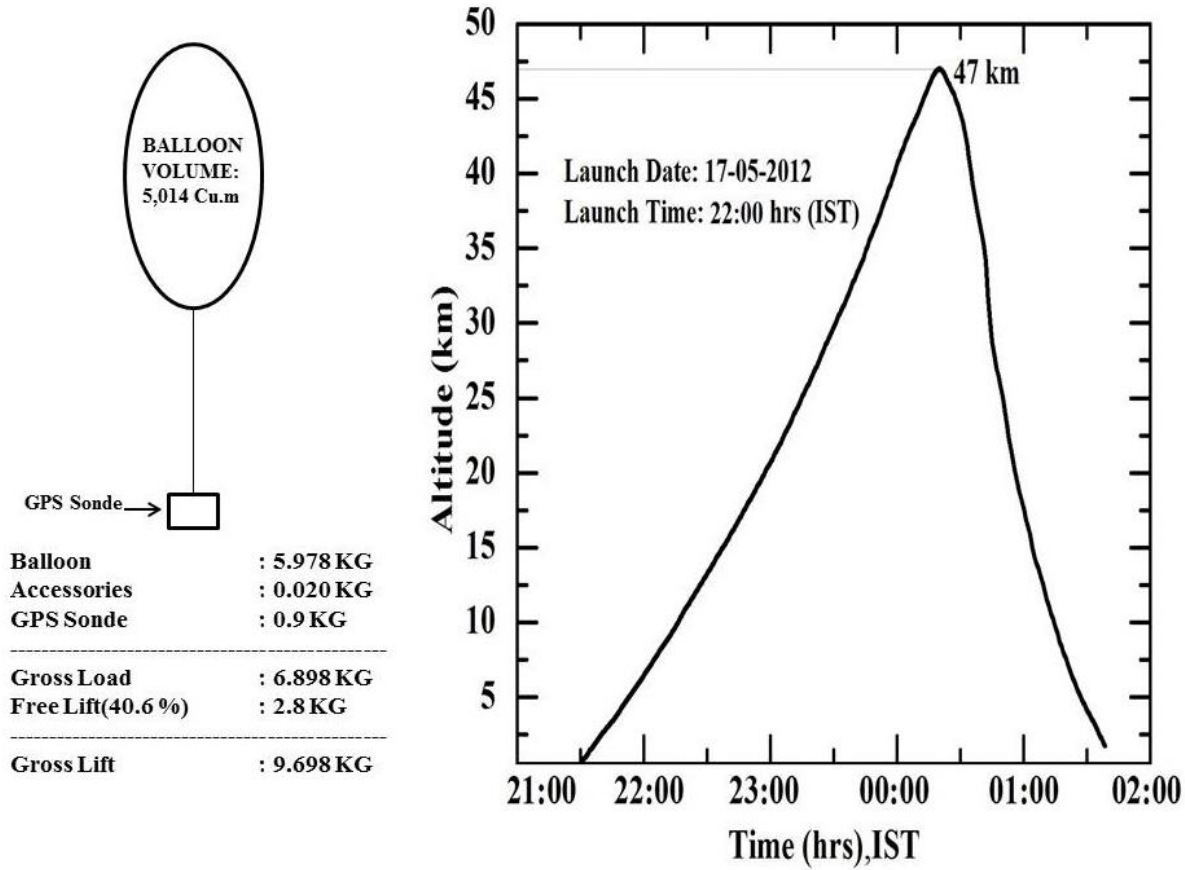

**Fig. 9.** The flight configuration and time-height curve for TS-16312 launch.

### 3.4. *Launching of HAA-02 balloon*

On 2012 May 29, the second HAA flight was conducted using 61,013 cu.m balloon (serial no: MS-16512; HAA-02).The flight accessories comprised 2 GPS-sondes, mini-timer and 2 plastic parachutes (total suspended load of 5.8 kg). This time, mini-timer was used to fire a single rope pyro cutter for flight termination, instead of Ni-chrome wire. It was set to operate at 08:00 hrs (IST) in order to allow sufficient time for the balloon to reach its maximum altitude. Two GPS-sondes were used to provide redundancy to GPS. Permission was required from ATC authorities for conducting balloon flight without using the ATC transponder and after obtaining the same, the balloon was filled with a gross lift of 47.8 kg including a free lift of 18% of gas (the free lift was reduced by 2% for realising higher burst altitude). The balloon was launched at 23:54 hrs (IST) with manual payload release. The balloon ascended with an average of 197 meters per min reaching an altitude of 39 km in 3 hrs 28 minutes, began to float at this altitude for 20 minutes, ascended again up to 42 km and continued to float at this altitude till 05:37 hrs (IST). After the sunrise, due to solar radiation, the balloon ascended rapidly and reached an altitude of 47.9 km at 06:02 hrs (IST) and then it burst due to build-up of super pressure and descended down on its own. The time-height curve shows the complete ascent, descent and float phases in Fig. 10.

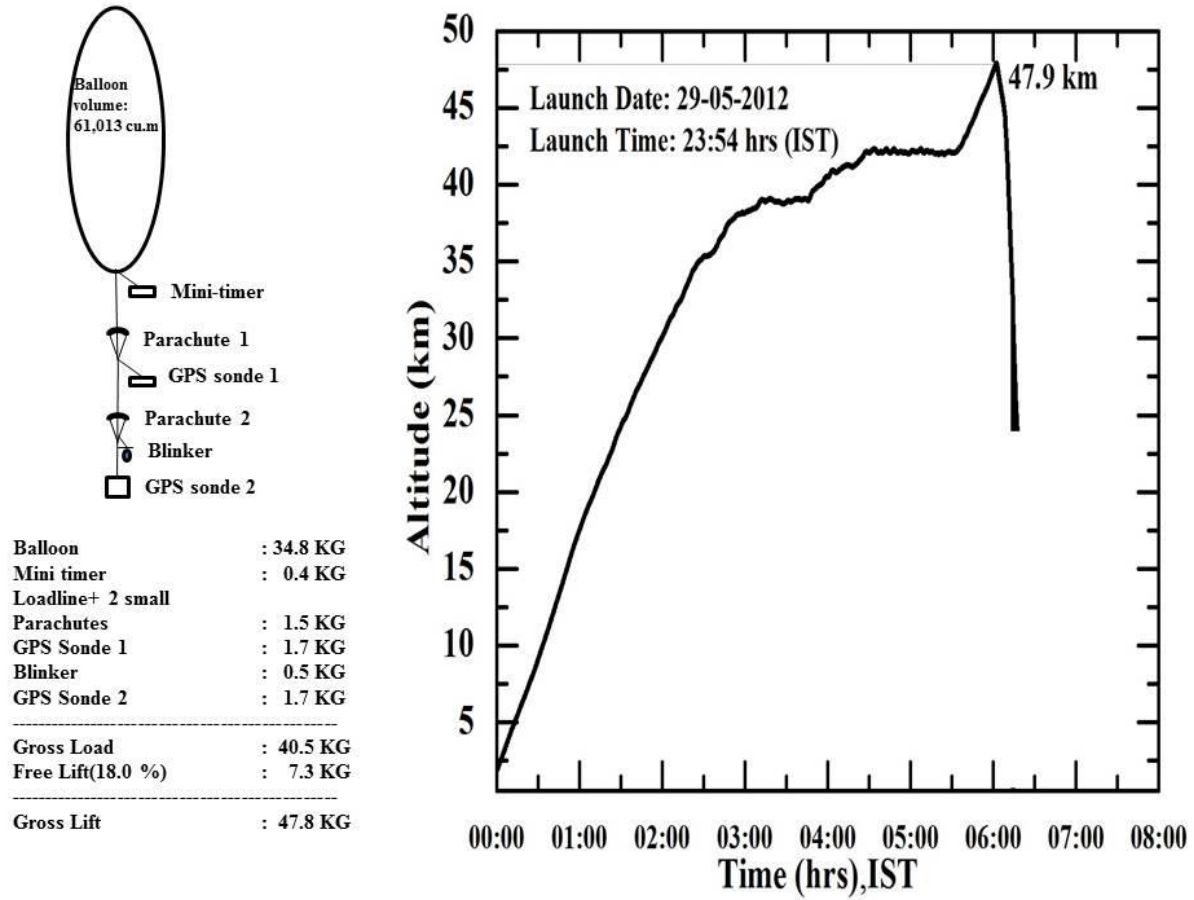

**Fig.10.** The flight configuration and time-height curve for HAA-02 balloon.

## 4. Flight analysis of high altitude balloons to Mesosphere

The analysis of previous balloon flights HAA-01 and HAA-02 showed that the balloons which were designed and fabricated to reach a theoretical float altitude of 52 km could reach only a maximum altitude of 48 km. A close look at the time-height curves of both flights revealed that both the balloons ascended up to 39 km and started floating for some time and then slowly reached their maximum altitudes after sunrise due to solar radiation input. It was felt that a low percentage free lift could have caused this phenomenon. An empirical formula to calculate the quantum of free lift based on the desired ascent rate as given below is being used at balloon facility (Morris, *et al.*, (1975) and Toyoo Abe, *et al.*, (2009)).

$$R = (f)^{1/2} / (g)^{1/3} \times K \qquad (1)$$

Where, R = rate of ascent (in meters/min),
    f = free lift (in kg), g = gross lift (in kg),
    K = constant (taken as 430 for rubber balloon ascents and 320 for plastic sounding balloons, these K values were arrived from the analysis of 10 years of balloon ascent rates at one hour Interval).

Both the HAA-01 and HAA-02 balloons were filled with 20% and 18% free lift corresponding to average ascent rates of 235 and 197 meters per min respectively. From the experience gained from these two flights, it was decided to increase the free lift percentage for future high altitude balloon flights and also schedule the launch close to sunrise time so that the balloon derives its additional lift from the solar radiation. Three more balloons were fabricated with volume same as HAA-01 and HAA-02, using film of uniform thickness for achieving further reduction in balloon weight. The details of balloons are given in Table 3.

Table 3. List of balloons fabricated with 3.8 µm film (2012 lot).

| S.No. | Balloon serial number | Weight (kg) |
|---|---|---|
| 1 | MS-16912 | 29.5 |
| 2 | MS-17012 | 33.5 |
| 3 | MS-17112 | 31.0 |

### 4.1. *HAA-03 balloon flight*

On 2014 January 7, a balloon launch was conducted using 61,013 cu.m volume high altitude balloon (serial no: MS-17112; HAA-03) fabricated out of 3.8 µm film weighing 31 kg. For this launch, a roller was made out of aluminium hollow tube and fitted onto a trolley, to hold the balloon bubble during inflation. Static launch method was adopted for this launch wherein the bottom part of the load line (payload) is tied to a vehicle and payload is suspended in the air with a tow balloon (launch balloon) of 20 kg lift capacity. The payload comprised single card command with an integrated timer, two new light weight i-Met-1 GPS-sondes, TSLM, ATC transponder, blinker and GSM-GPS unit. A handy cam is also incorporated for looking upwards to capture the balloon shape at various stages. Except GPS-sondes, all the units were packed together and the total instrument weight was 7.87 kg. A 4m parachute weighing 3 kg was used for safe payload landing. Thus the consolidated suspended weight was 10.87 kg. The free lift percentage realized for this balloon flight was 32%. A total gross lift of 55.27 kg of hydrogen gas was filled into the balloon in 35 minutes using P&T method. The surface wind was calm during inflation with a steady direction of $135^0$ (south-easterly). The balloon was released at 04:02 hrs (IST) from the roller and the payload was released by cutting the anchor nylon rope tied between payload and vehicle. The payload was released once the balloon, load line and parachute were air borne. Immediately after releasing the payload, telecommand was given from control room for separating the tow balloons, otherwise the tow balloon tied to payload would move vertically faster than main balloon because of the free lift percentage of hydrogen gas filled in the main balloon being less than the lift gas filled in the tow balloon. After one minute, the tow balloon was detached successfully and the total balloon and payload train became vertical. The schematic representation of static launch method is shown in Fig. 11.

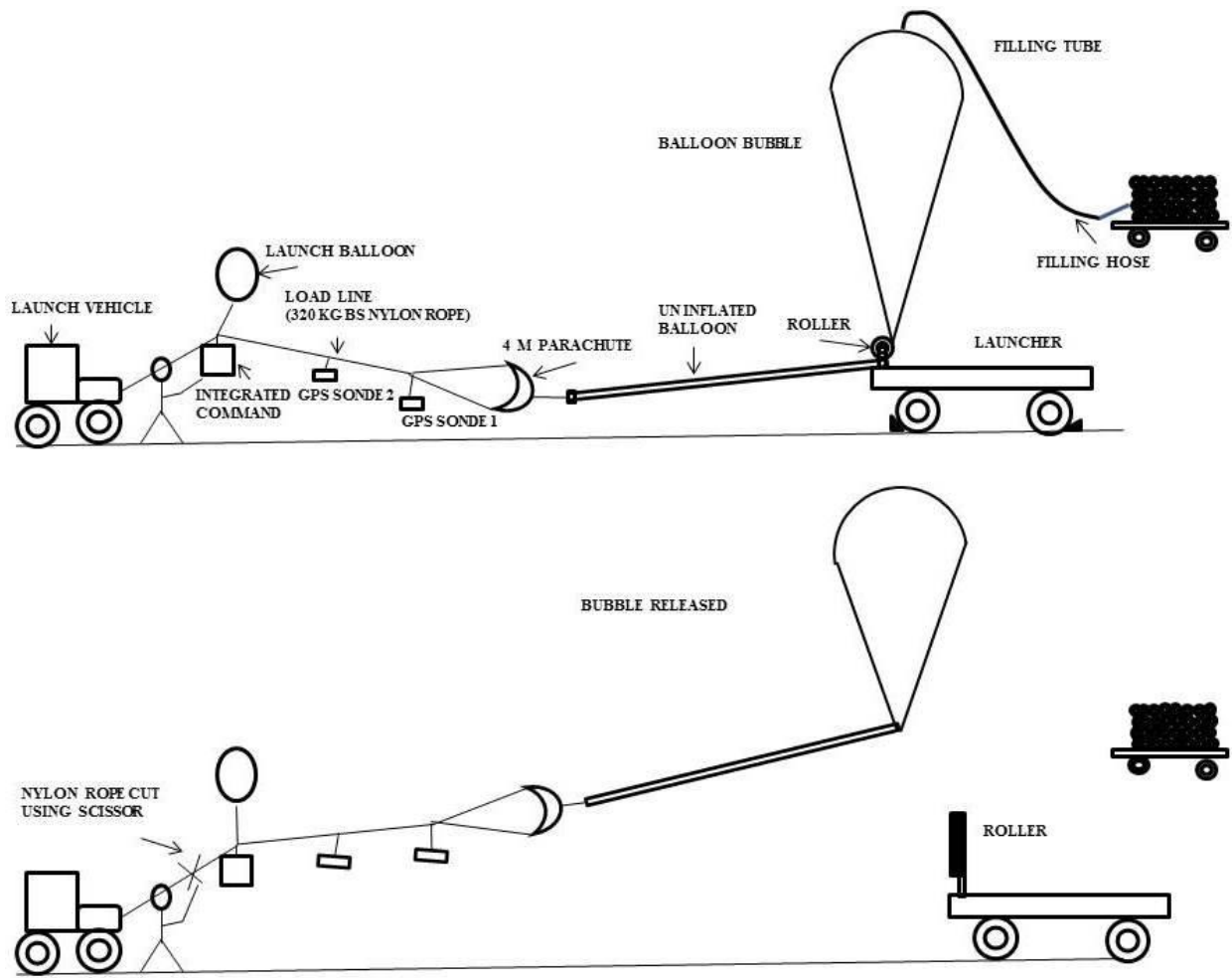

**Fig. 11.** Schematic representations of static launch method: Balloon bubble inflation (above) and balloon release from roller (below).

The balloon initially ascended at an average rate of 256 meters per min for 10 minutes and then accelerated at a faster rate of 590 meters per min in the next 10 minute range. An average ascent rate of 325 meters per min was observed during first hour. The balloon reached a maximum altitude of 51.66 km in 2 hrs 9 minutes with an average ascent rate of 396 meters per min and then the balloon started coming down slowly due to development of a small hole caused by super pressure. The balloon slowly descended to 4.7 km in 50 minutes and the flight termination command was given at 07:02 hrs (IST). The payload landed about 100 km north-east with respect to launching site. All the electronic packages functioned well during all the phases. An additional GSM-GPS unit used to track and guide the recovery team vehicle proved to be quite useful. The flight was successful in all respects. The balloon designed to reach 52 km performed well and reached its maximum altitude of 51.66 km carrying 10.9 kg payload. The flight configuration and the time-height curve are shown in Fig. 12 for the HAA-03 flight.

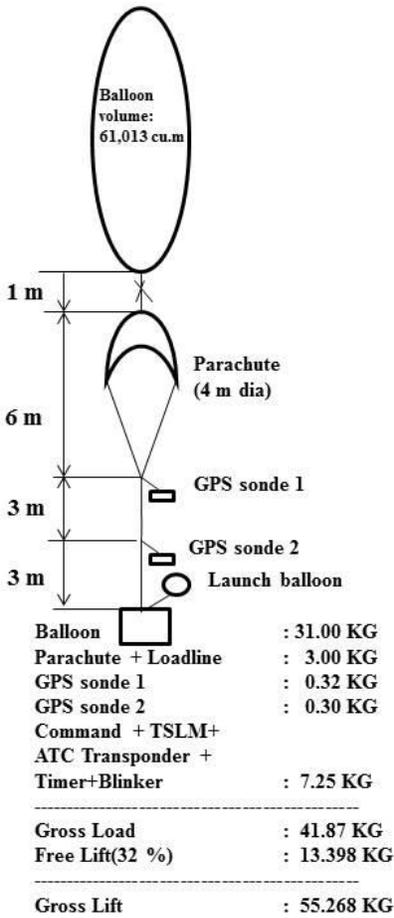
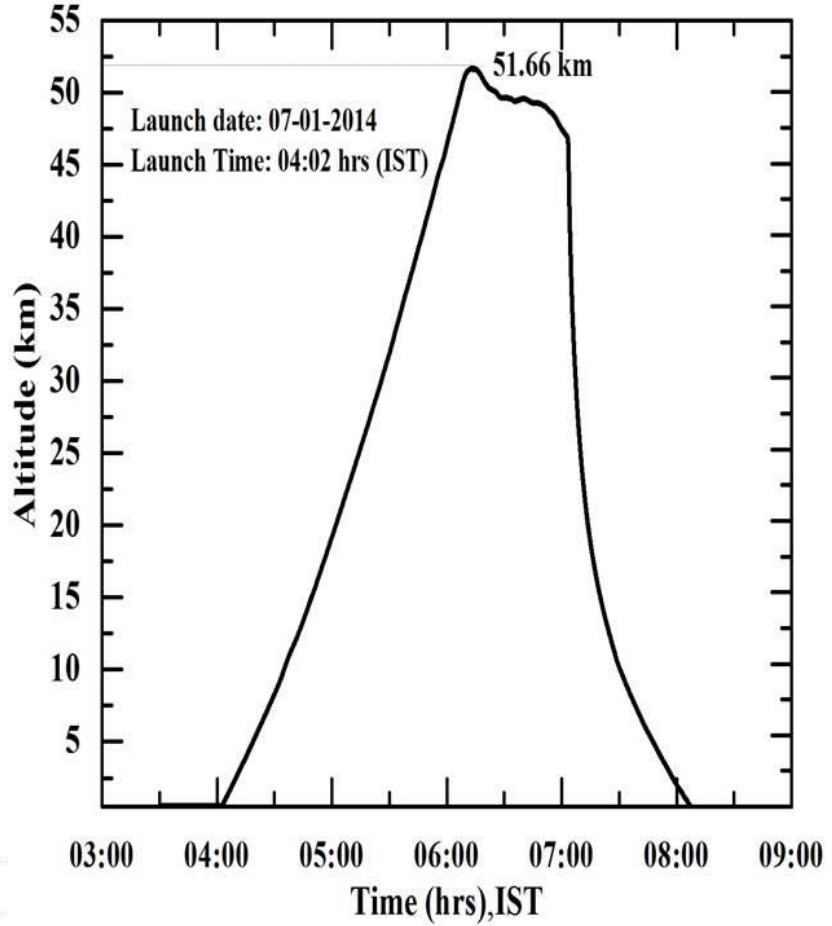

**Fig.12.** The flight configuration and time-height curve for HAA-03 balloon.

This is for the first time a balloon fabricated at balloon facility, Hyderabad penetrated into mesosphere in India. A small incremental change in free lift percentage worked here to push the balloon up to mesosphere. After recovery, all packages were found to be in good condition. So far only balloons from USA and Japan have achieved this feat. In 1972, USA launched a 1,350,000 cu.m volume winzen balloon and it reached to 51.8 km (http://www.extremespaceadventures.com/projecthubar). In 2004, Yamagami *et al.,* (2004) developed and fabricated a balloon of volume 60,000 cu.m with 3.4 μm ultra-thin balloon film and it reached a maximum altitude of 53 km. So far this is the highest altitude record reached by any scientific balloon in the world.

### 4.2. *HAA-04 balloon flight*

Encouraged by the results obtained in previous balloon flights, one more flight was conducted on 2014 January 21 with the same flight configuration as HAA-03 (serial no: MS: 17012; HAA-04). The balloon weight was slightly more than the HAA-03 balloon (33.5 kg). In this flight, the free lift percentage was fine-tuned by reducing it by 2% based on the experience of HAA-03 flight to achieve optimum ascent rate. The balloon operations were smooth and the surface wind conditions were also favourable with wind speed of 0.7 knots with a steady wind direction of $140^0$ (south-easterly). Total gross lift of 57.4 kg including 30% free lift was filled in the balloon using P&T method and filling process completed in 25 minutes. The same static launch method described earlier was adopted and the balloon was launched at 04:10 hrs (IST). The tow balloon was detached after 30 seconds using telecommand. The balloon ascended with an average ascent rate of 256 meters per min in 10 minutes and increased slightly for next 10 minutes at 287 meters per min. The balloon reached a maximum altitude of 51.83 km in 2 hrs 11 minutes with an average ascent rate of 396 meters per min and started coming down. The flight termination command was given at 06:32 hrs (IST) and the payload landed 31 km north of launch site. All the electronic flight packages functioned well during entire flight and recovered well from the impact point. The flight configuration and time-height curve are shown in Fig. 13. This balloon reached a maximum altitude higher than the HAA-03 balloon by successfully entering the mesosphere.

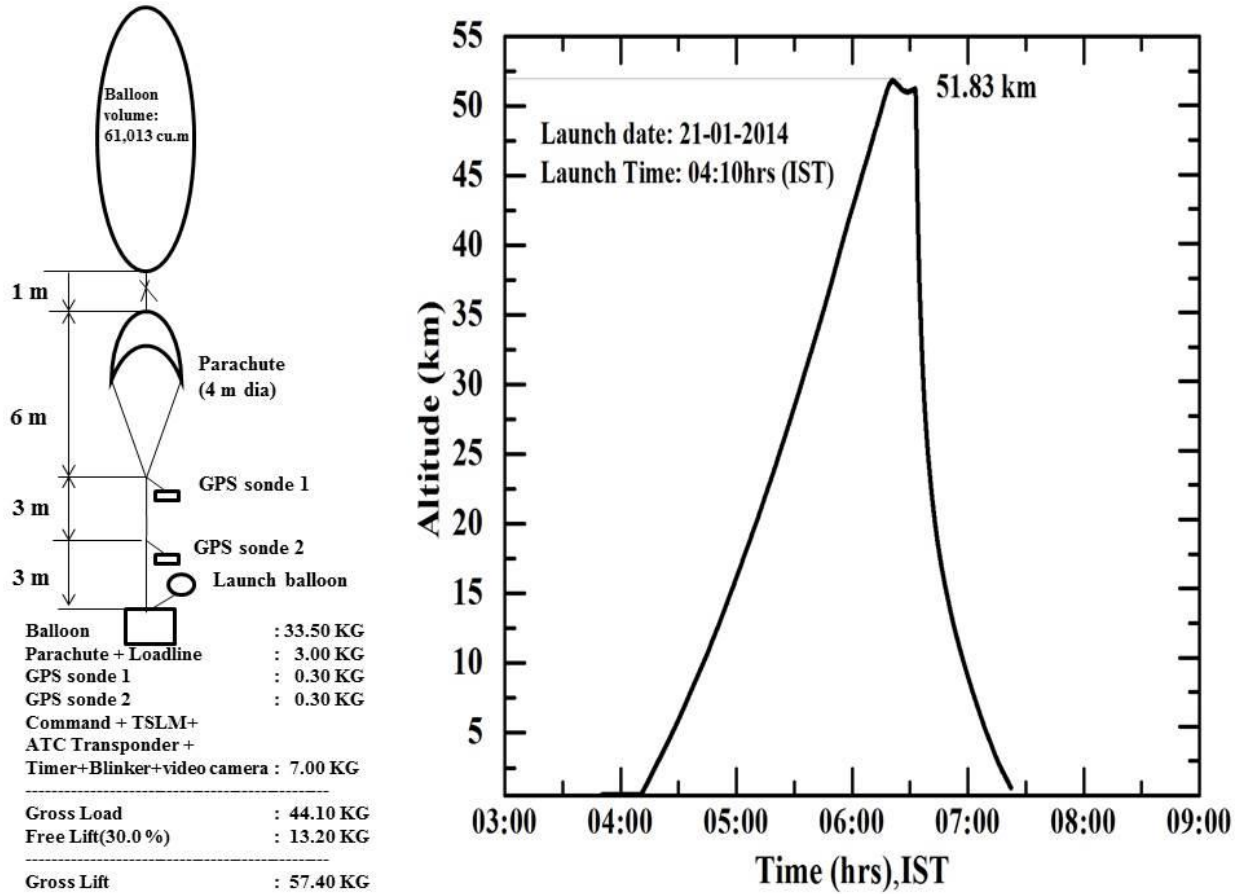

**Fig. 13.** The flight configuration and time-height curve for HAA-04 balloon.

### 4.3. *HAA-05 balloon flight*

Another high altitude ascent balloon launch was conducted on 2014 January 23 at 03:59 hrs (IST) using 61,013 cu.m balloon (serial no: MS-16612; HAA-05) and the balloon weight was 35.2 kg. The gross lift of 58.78 kg including 30% free lift was filled in the balloon using P&T method and filling process completed in 25 minutes. The balloon was launched using static launch method. The balloon carried a 7.0 kg payload consisting of telemetry encoder, s-band transmitter, single card command, ATC transponder, GSM-GPS mobile tracking unit and up looking video camera. The integrated package with a video camera and the fully inflated balloon photograph captured from the video camera at ceiling altitude are shown in Fig. 14.

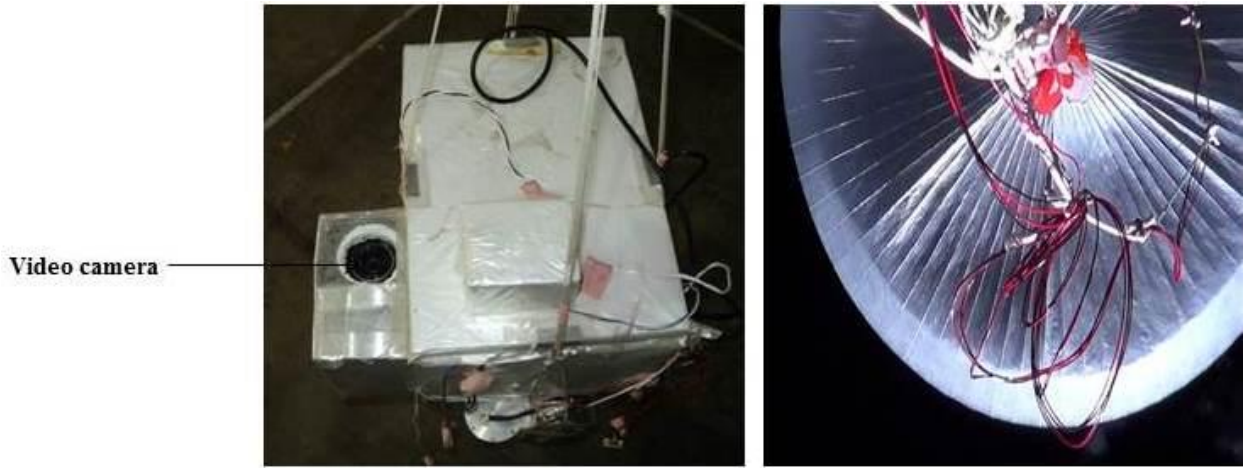

**Fig. 14.** The integrated package with video camera (left) and the fully inflated balloon photograph captured from the video camera at ceiling altitude (right).

The balloon reached a maximum altitude of 51.12 km at 06:02 hrs (IST) in 2 hrs 3 minutes with an average ascent rate of 419 meters per min. The balloon being a closed system burst after attaining the maximum altitude at 06:02 hrs (IST). For safe recovery of instruments, the payload separation mechanism was activated at 06:10 hrs (IST). The payload was recovered at 30 km north-east of the launch site. The flight configuration and time-height curves are shown in Fig. 15.

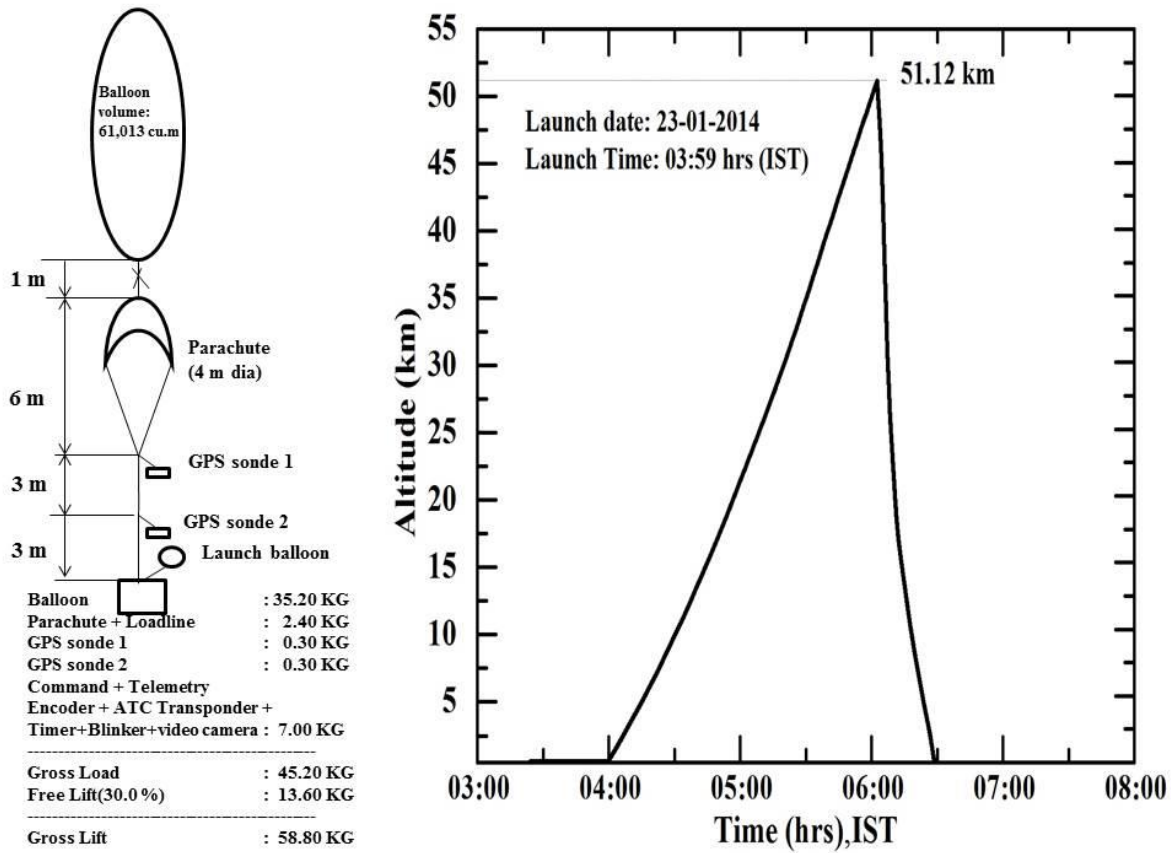

**Fig. 15.** The flight configuration and time-height curve for HAA-05 balloon.

## 5. Conclusions

A review of the project was carried out at this stage to consolidate the accomplishments and to chart out the future course of action based on the flight results. On the balloon side, the success achieved in terms of penetration of mesosphere within a time frame of three years has been very encouraging. However, there is a vast scope for reaching the world record if not breaking it exists. This can be accomplished by further reduction of the balloon film thickness without compromising the mechanical properties and making the balloon lighter. Table 4 shows the summary of all high altitude ascents.

Table 4. Summary of high altitude ascent balloon flights.

| S.No. | Balloon serial number | Launch date | Balloon Volume (Cu.m) | Balloon Weight, kg | Total Suspended weight, kg | Maximum altitude reached, km |
|---|---|---|---|---|---|---|
| 1 | TS-16111 | 01-12-2011 | 4,077 | 7.007 | 0.74 | 44.05 |
| 3 | TS-16212 | 08-03-2012 | 5,014 | 6.3 | 0.74 | 45.03 |
| 4 | HAA-01 | 25-04-2012 | 61,013 | 36.0 | 9.7 | 48.0 |
| 5 | TS-16312 | 17-05-2012 | 5,014 | 5.978 | 0.92 | 47.0 |
| 6 | HAA-02 | 29-05-2012 | 61,013 | 34.8 | 5.8 | 47.9 |
| 7 | HAA-03 | 07-01-2014 | 61,013 | 31.0 | 10.87 | 51.66 |
| 8 | HAA-04 | 21-01-2014 | 61,013 | 33.5 | 10.6 | 51.83 |
| 9 | HAA-05 | 23-01-2014 | 61,013 | 35.2 | 10.0 | 51.12 |

Analysis of the balloon weight of the five HAA flights showed that the weight drastically varied from a low of 31 kg to a high of 36 kg for the same volume of 61,013 cu.m . In order to reach maximum altitude, it is necessary to keep the weight of the balloons as low as possible. One major reason for this is the variation in the gauge thickness of the balloon film. It is relevant to mention here that for the manufacture of the film no improvements were made to the extrusion setup despite the need. It is strongly felt that there is ample scope for realising uniformity in gauge thickness by carrying out the following augmentation work.

1. Incorporation of a very accurate non-contact thickness profiler (preferably online) for dynamic control of the film thickness.
2. Providing air-conditioned environment which is an essential pre-requisite for extrusion of thin films.
3. Using a larger annular die of diameter 400 mm instead of the present annular die of diameter 300 mm.
4. Ensuring that the die gap is uniform (the present 300 mm annular die had a non-uniform die gap due to wear and tear).
5. Protecting the extrusion bubble from air draft using air curtains and ensuring its stability.

After implementing the above mentioned recommendations, it is possible for further down gauging of the balloon film to 2.8 µm, there by leading to significant reduction in balloon weight. On the flight support instrumentation side, the electronics group had done a commendable work in bringing the total balloon support instrumentation weight down to 7 kg. There is further scope for reduction of the system weight down to 3 kg. Once the goals of the balloon and electronics groups are accomplished, it will be possible to augment the payload carrying capability to 20 kg including flight support instrumentation.

**Acknowledgement**

We gratefully acknowledge the Tata Institute of Fundamental Research and the Balloon Facility Committee for constant encouragement and support for conducting the high altitude balloon experiments. We would like to express our special thanks to Professor R. K. Manchanda for initiating the development of ultra-thin film. We are also thankful to the staff members of electronics group for reducing weight of the electronic packages, the workshop group for developing a new mechanical launcher and modified roller, the balloon group for fabrication and launch of these thin film balloons and all other staff members of the TIFR Balloon Facility, Hyderabad for their help in conducting these high altitude balloon experiments.